\documentclass[12pt]{iopart}
\usepackage{graphicx}% Include figure files
  
\begin{document}

\title{Valence and magnetic instabilities in Sm compounds at high pressures}

\author{A Barla\dag\ddag, J-P Sanchez\dag, J Derr\dag, B Salce\dag, G Lapertot\dag, J Flouquet\dag, B P Doyle\ddag\S, O Leupold\ddag, R R{\"u}ffer\ddag, M M Abd-Elmeguid$\|$ and R Lengsdorf$\|$}

\address{\dag\ D\'epartement de Recherche Fondamentale sur la Mati\`ere Condens\'ee, CEA Grenoble, 17 rue des Martyrs, F-38054 Grenoble Cedex 9, France}

\address{\ddag\ European Synchrotron Radiation Facility, 6 rue Jules Horowitz, F-38043 Grenoble Cedex 9, France}

\address{\S\ Laboratorio TASC-INFM, Area Science Park, Basovizza S.S. 14 Km 163.5, I-34012 Trieste, Italy}

\address{$\|$\ II. Physikalisches Institut, Universit{\"a}t zu K{\"o}ln, Z{\"u}lpicher Strasse 77, D-50937 K{\"o}ln, Germany}

\ead{barla@drfmc.ceng.cea.fr}

\begin{abstract}
We report on the study of the response to high pressures of the electronic and magnetic properties of several Sm-based compounds, which span at ambient pressure the whole range of stable charge states between the divalent and the trivalent. Our nuclear forward scattering of synchrotron radiation and specific heat investigations show that in both golden SmS and SmB$_{6}$ the pressure-induced insulator to metal transitions (at 2 and $\sim$~4~-~7~GPa, respectively) are associated with the onset of long-range magnetic order, stable up to at least 19 and 26~GPa, respectively. This long-range magnetic order, which is characteristic of Sm$^{3+}$, appears already for a Sm valence near 2.7. Contrary to these compounds, metallic Sm, which is trivalent at ambient pressure, undergoes a series of pressure-induced structural phase transitions which are associated with a progressive decrease of the ordered 4\textit{f} moment.
\end{abstract}


\submitto{\JPCM}

\maketitle

\section{Introduction}

Considerable interest has been devoted in the past decades to the study of intermetallic compounds based on anomalous lanthanides (Ce, Yb, Tm, Eu and Sm) because of the rich variety of properties they have (intermediate valence, heavy fermion behaviour, $\dots$). More recently, particular attention was focused on 4\textit{f} intermetallics which are close to a magnetic instability, because of the observation of unconventional superconductivity in the vicinity of a quantum critical point (QCP), where the magnetic order is destroyed \cite{MGJ98}. In the specific case of Sm, the competition between the two possible valence states, nonmagnetic Sm$^{2+}$ with a 4\textit{f}$^{6}$:$^{7}F_{0}$ configuration and magnetic Sm$^{3+}$ with a 4\textit{f}$^{5}$:$^{6}H_{5/2}$ configuration, can sometimes lead to the formation of an intermediate valent ground state which shows peculiar electronic and magnetic properties \cite{Wac94}. The charge state can be eventually tuned towards integer trivalency by the application of external pressure, with the consequent onset of long-range magnetic order owing to the fact that Sm$^{3+}$ is a Kramer's ion. Well-known examples of intermediate valent Sm compounds are the high-pressure "golden" phases of the Sm monochalcogenides (SmS, SmSe and SmTe) \cite{JNB70} and SmB$_{6}$ \cite{MBG69}. They belong to the class of Kondo insulators or narrow-gap semiconductors, which behave at high temperature like an array of independent localized moments interacting with itinerant conduction electrons, whereas at low temperature they develop clear narrow-gap properties. 

At ambient pressure SmS is a nonmagnetic semiconductor (black phase), with nearly divalent Sm ions, crystallizing in the NaCl-type structure. At a very low pressure $p_{\mathrm{B-G}}$~$\sim$~0.65~GPa at room temperature it undergoes an isostructural first order phase transition to an intermediate valent metallic state (golden phase), with a large volume collapse ($\sim$8\%) and a valence $v$~$\approx$~2.6 just above $p_{\mathrm{B-G}}$ \cite{Wac94}. The temperature dependence of the electrical resistivity shows the opening of a small gap at low temperatures for pressures between $p_{\mathrm{B-G}}$ and $p_{\mathrm{\Delta}}$~=~2~GPa. The gap closes continuously as pressure is increased and disappears for $p$~$>$~$p_{\mathrm{\Delta}}$, leaving the system in a metallic ground state \cite{HoW81,LRH81}. In the golden phase the valence increases continuously with pressure, with an inflection point at $p_{\mathrm{\Delta}}$ ($v$~$\sim$~2.7) \cite{RKK82}, and the trivalent state is reached only at considerably higher pressure ($p_{\mathrm{3+}}$~$>$~10~GPa) \cite{RKK82,DAR04}. Recently a first order transition at $p_{\mathrm{\Delta}}$ with the appearance of long-range magnetic order associated with intermediate valence has been evidenced by nuclear forward scattering (NFS) of synchrotron radiation \cite{BSH04} and specific heat \cite{HDB04}. The magnetically ordered state behaves like that of a stable trivalent compound, with values of the hyperfine parameters as expected for a $\Gamma_{8}$ ground state (produced by the action of a cubic crystal field on the Sm$^{3+}$ ions) and almost independent from pressure in the range 2~-~19~GPa, while the ordering temperature $T_{\mathrm{m}}$ increases smoothly from 15~K at $p_{\mathrm{\Delta}}$ to 24~K at 8~GPa. Considerable short-range magnetic correlations develop above $T_{\mathrm{m}}$ and persist up to about 2$T_{\mathrm{m}}$.

In analogy to golden SmS, SmB$_{6}$ is an intermediate valent ($v$~=~2.6 at ambient pressure and room temperature) insulator \cite{Wac94}. However at temperatures above $\sim$~70~K the electrical resistivity and the Hall effect are rather typical of a bad metal, and only for $T$~$<$~70~K the conductivity starts to decrease with temperature by several orders of magnitude (two to five, depending on the quality of the sample) because of the opening of an insulating gap of the order of 10~-~20~meV \cite{NWL71,ABW79}. Despite the intermediate valent ground state of SmB$_{6}$, the form factor as determined by neutron scattering is that expected for Sm$^{2+}$ \cite{BAG95}, as already observed for similar systems like golden SmS and TmSe \cite{BCC81}. However the magnetic excitation spectrum shows that the "true" ground state of SmB$_{6}$ is an unusual quantum mechanical superposition of a divalent component of 4\textit{f}$^{6}$ configuration with a 4\textit{f}$^{5}$5\textit{d} loosely bound state, where a 4\textit{f} electron is only partially delocalised and free to move within a limited volume \cite{AMR95,MiA95,KiM95}. Although this ground state is nonmagnetic, $^{11}$B NMR measurements have evidenced the onset of dynamical magnetic correlations below $\sim$~15~K \cite{TYK81}. When external pressure is applied, the semiconducting gap decreases and finally closes at a critical pressure $p_{\mathrm{\Delta}}$~$\approx$~4~-~7~GPa, above which the system behaves as a metal at all temperatures \cite{BMW83,MBB85,CAF95,GBB03}. At $p_{\mathrm{\Delta}}$ the resistivity has a less than quadratic dependence on temperature down to 1.5~K, suggesting that a QCP is crossed when the gap closes \cite{GBB03}. The valence has been shown to increase smoothly from 2.6 at ambient pressure to 2.8 at 10~GPa, with an inflection point at $\sim$~5.5~GPa at the gap closure \cite{BKK90,Roh87}, in analogy with SmS at 2~GPa.

Contrary to SmS and SmB$_{6}$, metallic Sm is at ambient pressure in a trivalent state with long-range antiferromagnetic order below $\sim$~106~K. However, recent theoretical calculations predict that the 4\textit{f} electrons should become completely delocalized with a volume contraction corresponding to a pressure of the order of $\sim$~100~GPa \cite{SEW93}, which can be achieved with the diamond anvil cell (DAC) technique. Sm crystallizes at ambient pressure in the Sm-type structure, with a unit cell consisting of the periodic stacking of nine hexagonal planes (A$_{c}$B$_{h}$A$_{h}$B$_{c}$C$_{h}$B$_{h}$C$_{c}$A$_{h}$C$_{h}$ $\dots$) \cite{Daa54}. The sublattice formed by the six planes with hexagonal nearest neighbour environment (indicated by the subscript $h$) orders antiferromagnetically at $T_{\mathrm{N}h}$~=~106~K, while the planes with approximately cubic local symmetry (indicated by the subscript $c$) order independently only below $T_{\mathrm{N}c}$~=~14~K \cite{KoM72,MoK73}. The 4\textit{f} ordered magnetic moment is $\sim$~0.6~$\mu_{\mathrm{B}}$ \cite{MoK73}, lower than the free ion value of 0.71~$\mu_{\mathrm{B}}$ for the Sm$^{3+}$ ion. High pressure resistivity studies performed up to 43~GPa show that, in the range between ambient pressure and 6~GPa, $T_{\mathrm{N}h}$ decreases while $T_{\mathrm{N}c}$ increases with pressure \cite{DLD87}. Above 6~GPa the two N\'eel transitions occur at the same temperature, increasing with pressure and reaching $\sim$~135~K at 43~GPa. At higher pressure, only structural investigations have been performed, showing the presence of five successive structural phase transitions between ambient pressure and 189~GPa \cite{VAW91,ZPH94}. In particular above 91~GPa Sm adopts a body-centred tetragonal structure, in which calculations show that the 4\textit{f} electrons are delocalized and that Sm is an itinerant magnet with spin moment of $\geq$~4~$\mu_{\mathrm{B}}$ \cite{SEW93}. This moment is very sensitive to pressure and disappears when the volume is sufficiently reduced.

In this Article we present further results on golden SmS (a detailed discussion of previous recent results is given in references~\cite{BSH04,HDB04}) and new results on SmB$_{6}$ and metallic Sm at high pressures. These results were obtained by combining $^{149}$Sm nuclear forward scattering (NFS) of synchrotron radiation with high pressure specific heat measurements. From the NFS data we obtained information about the magnetic hyperfine field at the $^{149}$Sm nuclei as a function of pressure and temperature, and from the combination of NFS and specific heat we could determine the ordering temperature $T_{\mathrm{m}}$ of the Sm atoms and its pressure dependence. In particular we have obtained clear evidence that SmB$_{6}$, in analogy with golden SmS, develops long-range magnetic order at pressures above $p_{\mathrm{\Delta}}$. These magnetically ordered states are stable up to at least 19 and 26~GPa for SmS and SmB$_{6}$, respectively. Moreover, in the low pressure intermediate valent state of both compounds (and down to ambient pressure for SmB$_{6}$) short-range magnetic correlations are present at low temperatures. In the case of metallic Sm, the magnetic hyperfine field decreases monotonically with pressure and is reduced by almost 40~\% at 46~GPa with respect to ambient pressure.

\section{Experimental methods}

The SmS sample was prepared as described in reference~\cite{BSH04}. Samarium hexaboride single crystals were grown by a standard aluminum flux technique \cite{CaF92,FiR89}. SmB$_{6}$ powder was pre-synthesized by reacting boron on samarium oxyde (borothermal reduction) under high vacuum using RF heating. The quality of the sample was checked by x-ray powder diffraction and showed the presence of pure SmB$_{6}$ with no trace of parasitic phases. For both SmS and SmB$_{6}$ the $^{149}$Sm NFS measurements have been performed on powders made with isotopically enriched (to 97~\%) Sm, while the specific heat measurements have been performed on single crystals made with natural Sm. The metallic Sm sample was a polycrystalline foil, isotopically enriched to 97~\% in $^{149}$Sm, commercially available from Oak Ridge National Laboratories.

High pressure was applied to the samples using the diamond anvil cell technique, with argon or nitrogen used as pressure transmitting media. The pressure was measured and changed always at room temperature for the NFS measurements, whereas it was measured and changed in situ using the setup described in reference~\cite{STD00} for the specific heat measurements performed at the CEA Grenoble. The $^{149}$Sm NFS measurements (resonant energy $E_{0}$~=~22.494~keV; 5/2~-~7/2 transition) were performed at the undulator beamline ID22N \cite{RuC96} of the European Synchrotron Radiation Facility, Grenoble, France, and a more detailed description of the experimental setup is given in reference~\cite{BSH04}. NFS is a technique related to the M\"ossbauer effect, thus similar microscopic information to that inferred from conventional M\"ossbauer spectroscopy can be obtained. This allows one to determine the pressure dependence of the magnetic hyperfine field $B_{\mathrm{hf}}$, of the ordering temperature $T_{\mathrm{m}}$ and of the electric field gradient (EFG) $V_{\mathrm{zz}}$ at the $^{149}$Sm nuclei. The change of the EFG with $p$ is obtained from the pressure dependence of the electric quadrupole splitting $\Delta$$E_{\mathrm{Q}}$~=~$e$$V_{\mathrm{zz}}$$Q_{\mathrm{g}}$ where $Q_{\mathrm{g}}$ is the nuclear quadrupole moment of the $I_{\mathrm{g}}$~=~7/2 ground state.

\section{Pressure induced magnetic order in SmS and SmB$_{6}$}

\subsection{Long-range magnetic order}

\subsubsection{Ground state properties as seen by NFS}

\begin{figure}[tbh]
\begin{center}
\includegraphics[scale=0.5,clip=]{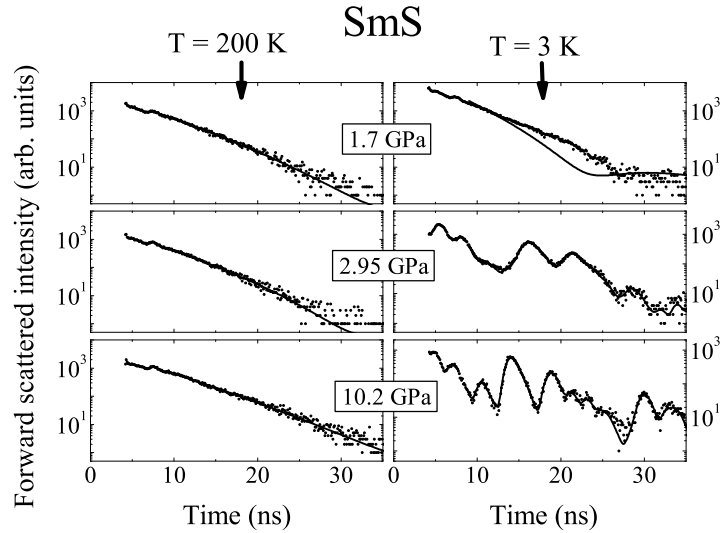}
\end{center}
\caption{\label{fig:one} $^{149}$Sm NFS spectra of SmS at $T$~=~200 and 3~K for some selected pressures. The dots represent experimental data points, while the lines are fits.}
\end{figure}
\begin{figure}[tbh]
\begin{center}
\includegraphics[scale=0.5,clip=]{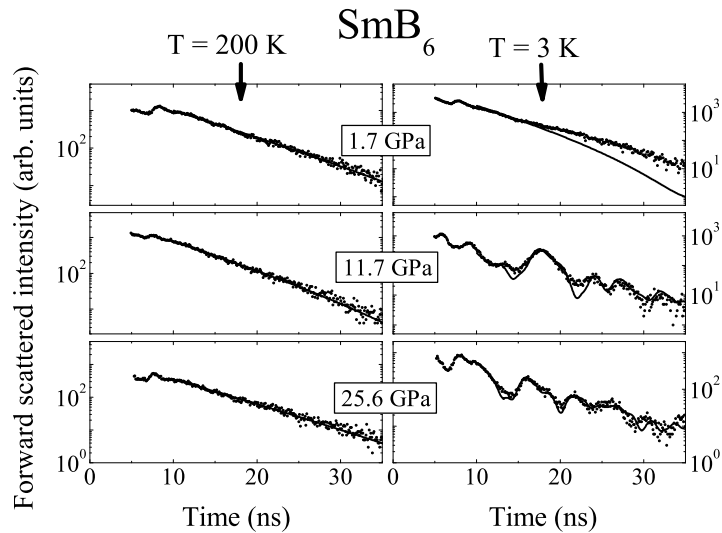}
\end{center}
\caption{\label{fig:two} $^{149}$Sm NFS spectra of SmB$_{6}$ at $T$~=~200 and 3~K for some selected pressures. The dots represent experimental data points, while the lines are fits.}
\end{figure}

Figures~\ref{fig:one} and \ref{fig:two} show typical NFS spectra recorded at various temperatures and pressures for golden SmS and SmB$_{6}$, respectively. At high temperatures (left panel in the two figures) and for all pressures the spectra are characteristic of unsplit nuclear levels, as expected for Sm ions in the absence of magnetic order and in an environment of cubic symmetry (NaCl-type stucture for SmS and CsCl-type structure for SmB$_{6}$). A clear quantum beat pattern, due to the combined action of magnetic dipole and electric quadrupole interactions on the nuclear levels of $^{149}$Sm, appears at low temperatures (right panel of the two figures, $T$~=~3~K) for pressures higher than $p_{\mathrm{c,SmS}}$~$\approx$~2~GPa for SmS and $p_{\mathrm{c,SmB_{6}}}$~$\approx$~9~GPa for SmB$_{6}$. This is a clear indication that magnetic order sets in in both compounds at high pressure and low temperature. The analysis of the spectra, performed with the software MOTIF \cite{Shv99,Shv00} using the full dynamical theory of nuclear resonance scattering, including the diagonalization of the complete hyperfine Hamiltonian, reveals that for SmS at 2.35~GPa and 3~K only a fraction (corresponding to $\sim$~72~\%) of the Sm atoms shows magnetic order, the rest being paramagnetic. However this fraction increases rapidly with pressure and reaches 100~\% above $\sim$~3~GPa. At 2.35~GPa we deduce a value of 261(10)~T and -1.50(6)~mm/s for the magnetic hyperfine field $B_{\mathrm{hf}}$ and the induced quadrupole interaction $\Delta$$E_{\mathrm{Q}}$, respectively \cite{BSH04}. For SmB$_{6}$ at 9.7~GPa and 3~K the analysis reveals that all Sm ions are magnetically ordered. The saturation values of the hyperfine parameters at this pressure and 3~K are $B_{\mathrm{hf}}$~=~246(20)~T and $\Delta$$E_{\mathrm{Q}}$~=~-1.27(12)~mm/s. In the vicinity of the critical pressures $p_{\mathrm{c,SmS}}$ and $p_{\mathrm{c,SmB_{6}}}$ the low temperature NFS spectra cannot be accounted for by a single set of hyperfine parameters, but broad distributions of both $B_{\mathrm{hf}}$ and $\Delta$$E_{\mathrm{Q}}$ are present at the $^{149}$Sm nuclei, larger in SmB$_{6}$ than in SmS. The average values of the hyperfine parameters are considerably reduced with respect to the free ion values for Sm$^{3+}$ ($B_{\mathrm{hf}}$~=~338~T and $\Delta$$E_{\mathrm{Q}}$~=~-2.1~mm/s \cite{Ble72}). This can be ascribed to the effect of the cubic crystal field acting on the Sm$^{3+}$ ions: the lowest multiplet $^{6}H_{5/2}$ is split into a $\Gamma_{7}$ doublet and a $\Gamma_{8}$ quartet (here and in the following we will neglect the possible mixing of the ground multiplet with excited multiplets through exchange interactions), associated to completely different values of the hyperfine parameters ($B_{\mathrm{hf}}$~=~113~T and $\Delta$$E_{\mathrm{Q}}$~=~0 for the $\Gamma_{7}$, while $B_{\mathrm{hf}}$~=~250~T and $\Delta$$E_{\mathrm{Q}}$~=~-1.7~mm/s for the $\Gamma_{8}$). Our measured values for both compounds, especially those for the magnetic hyperfine field, compare well with those calculated for a $\Gamma_{8}$ state. However, one must keep in mind that, at least for the lowest pressures above $p_{\mathrm{c,SmS}}$ and $p_{\mathrm{c,SmB_{6}}}$, the magnetic order coexists with intermediate valence and therefore the ground state wavefunction can be considerably more complicated than the simple one describing the $\Gamma_{8}$ quartet. For both compounds the hyperfine parameters in the magnetically ordered state have only a very weak dependence upon pressure in the range studied, showing however a slight continuous increase of magnitude as pressure increases. At the same time, the widths of the distributions of $B_{\mathrm{hf}}$ and $\Delta$$E_{\mathrm{Q}}$ decrease with increasing pressure but remain finite even at the highest pressures (the distribution of $B_{\mathrm{hf}}$ has a relative width of $\sim$~3~\% at $p$~$>$~5~GPa for SmS and of $\sim$~7~\% at $p$~$>$~11~GPa for SmB$_{6}$). These facts prove that the pressure induced magnetic order in these two compounds is very stable and that the transition into the trivalent state (which occurs for $p$~$>$~10~-~12~GPa in the case of SmS \cite{DAR04}) is not associated to any anomaly in the pressure dependence of the hyperfine parameters. This fact and the good agreement between the measured values of $B_{\mathrm{hf}}$ and $\Delta$$E_{\mathrm{Q}}$ and those expected for Sm$^{3+}$ ions in a cubic crystal field might be an indication of the fact that, in the high pressure metallic state, these two compounds behave like stable trivalent compounds, independently of $v$ and similarly to TmSe above 3~GPa \cite{RFH80}.

\subsubsection{Temperature dependence: NFS and specific heat}

\begin{figure}[tbh]
\begin{center}
\includegraphics[scale=0.5,clip=]{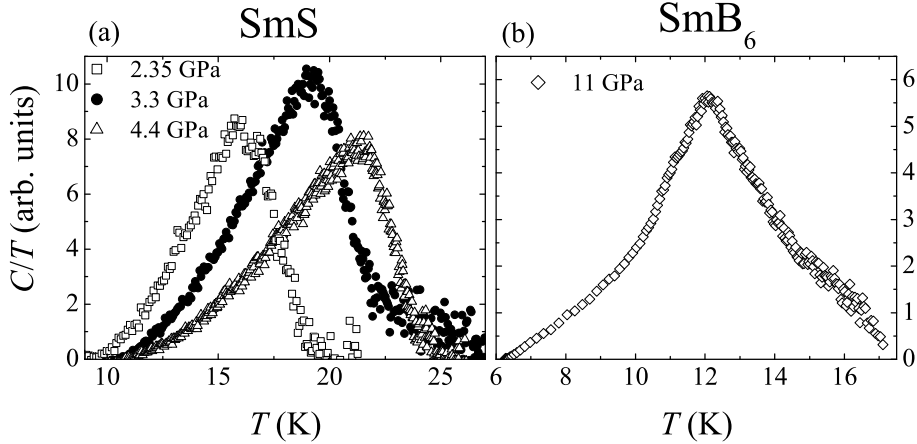}
\end{center}
\caption{\label{fig:three} The specific heat of (a) SmS and (b) SmB$_{6}$
 as a function of temperature and pressure.}
\end{figure}

In order to shed more light on the properties of the high-pressure magnetically ordered state, we have performed temperature dependent NFS and specific heat measurements at various pressures. The temperature and pressure dependences of the specific heat for golden SmS and SmB$_{6}$ are shown in figure~\ref{fig:three}(a) and (b), respectively. The presence of sharp anomalies at temperatures $T_{\mathrm{m}}(p)$ for pressures higher than 2~GPa for SmS and $\sim$~9~GPa for SmB$_{6}$ is a clear indication that the order observed by NFS in both systems at high pressure is long-range magnetic order. $T_{\mathrm{m}}$ is found to increase from 15~K at 2~GPa to 24~K at 8~GPa in SmS, whereas it has a value of $\sim$~12~K, independent from pressure up to 16~GPa, in SmB$_{6}$. 
\begin{figure}[tbh]
\begin{center}
\includegraphics[scale=0.4,clip=]{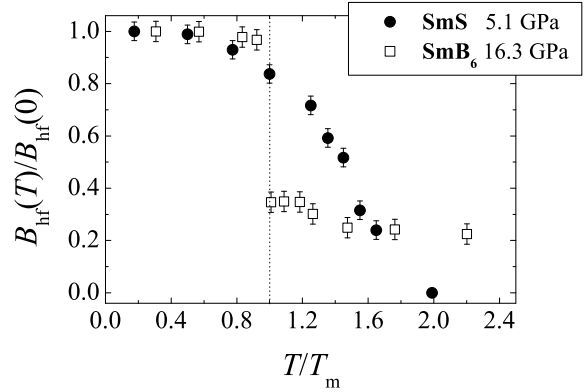}
\end{center}
\caption{\label{fig:four} The reduced magnetic hyperfine field of SmS (as an example at 5.1 GPa) and SmB$_{6}$ (as an example at 16.3 GPa) as a function of the reduced temperature. $T_{\mathrm{m}}$ is the temperature for the onset of long-range magnetic order as determined by specific heat.}
\end{figure}
The temperature dependence of the NFS spectra of SmS for $p$~$>$~2~GPa does not reveal any sharp transition at $T_{\mathrm{m}}$ as for what concerns the values of $B_{\mathrm{hf}}$ (see figure~\ref{fig:four}) and $\Delta$$E_{\mathrm{Q}}$, but a continuous decrease with increasing temperature up to $\sim$~2$T_{\mathrm{m}}$, where they become zero. The presence of hyperfine interactions at temperatures higher than $T_{\mathrm{m}}$ can be seen as due to the persistence of short-range magnetic correlations (probably of dynamical nature) in the paramagnetic phase. 
\begin{figure}[tbh]
\begin{center}
\includegraphics[scale=0.4,clip=]{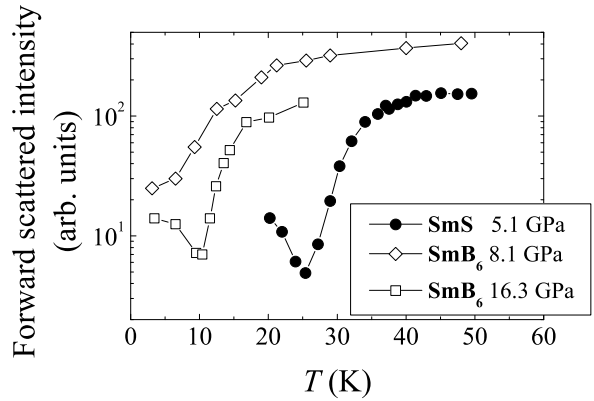}
\end{center}
\caption{\label{fig:five} The NFS intensity for SmS (as an example at 5.1 GPa) and SmB$_{6}$ (as an example at 8.1 and 16.3 GPa) as a function of temperature.}
\end{figure}
However, an anomaly in the temperature dependence of the NFS countrate (which is a complicated function of the physical properties of the compound under study) is clearly visible (see figure~\ref{fig:five}), with a minimum in the vicinity of $T_{\mathrm{m}}$ and the following steep increase upon entrance into the paramagnetic state. A similar behaviour is found in SmB$_{6}$, with a minimum in the countrate at the value of $T_{\mathrm{m}}$ determined by the specific heat measurements and the subsequent increase as temperature increases, as shown in figure~\ref{fig:five}. However, in this case a clear transition from the long-range magnetic order below $T_{\mathrm{m}}$ to a phase characterized by short-range magnetic correlations is directly observed in the NFS spectra, as shown in figure~\ref{fig:four}: above $T_{\mathrm{m}}$ the average value of $B_{\mathrm{hf}}$ and $\Delta$$E_{\mathrm{Q}}$ is considerably reduced, their distribution is much broader and finally only a fraction of the Sm ions feels these interactions, which slowly decrease and disappear at temperatures higher than $\sim$~50~K for all pressures up to 26~GPa.

\subsection{Short-range magnetic correlations}

In the top panels of figures~\ref{fig:one} and \ref{fig:two} spectra are shown which are characteristic of the low pressure ($p$~=~1.7~GPa) semiconducting intermediate valent states of golden SmS and SmB$_{6}$, respectively. At 3~K, the NFS spectra cannot be analyzed properly if the assumption is made that the nuclear levels are not split by any hyperfine interaction, as would be expected for a nonmagnetic ground state: this is clear from the fact that the best fit (shown in the figures as a line) to the two NFS spectra at 3~K and 1.7~GPa simply does not reproduce the measured data. The only reasonable agreement between calculated and experimental curves can be obtained in the hypothesis that at least a fraction of the Sm ions feels hyperfine interactions due to the presence of (short-range) magnetic correlations, which can be seen as a precursor of the incipient onset of long-range magnetic order above $p_{\mathrm{c,SmS}}$ or $p_{\mathrm{c,SmB_{6}}}$. Because of the extremely broad distributions of these hyperfine interactions, they do not show up as a clear beat pattern in the NFS spectra but rather only as a slowing down of the nuclear decay and prevent a precise analysis of their strength. The broad maxima present in the temperature dependence of the specific heat for pressures 1.3~$<$~$p$~$<$~2~GPa in the case of SmS and 7~$<$~$p$~$<$~9~GPa for SmB$_{6}$ can as well be interpreted in the framework of the onset of low temperature short-range slow magnetic correlations for pressures just below $p_{\mathrm{c,SmS}}$ and $p_{\mathrm{c,SmB_{6}}}$. Although the NFS spectra of SmB$_{6}$ are characterized by the presence of short-range correlations even at ambient pressure, appearing always below 50~-~100~K independently of pressure, two regimes can be distinguished in the temperature dependence of the NFS countrate: for 0~$\leq$~$p$~$\leq$~5~GPa there is no appreciable variation between 3 and 300~K, whereas for 5~$\leq$~$p$~$\leq$~9~GPa a steep increase is observed between $\sim$~10~K and $\sim$~50~K, similar to the one observed for higher pressures, but no minimum is present (see figure~\ref{fig:five}). The interval between 5 and 9~GPa, approximately where the semiconducting gap closes, can therefore be seen as a region of transition between the low-pressure gapped state and the high-pressure state with long-range magnetic order. 

The ground state of intermediate valent golden SmS and of ambient-pressure SmB$_{6}$ is generally considered as a nonmagnetic state: the magnetic susceptibility does not show any paramagnetic Curie-Weiss divergence at low temperatures, as would be expected for Sm$^{3+}$ ions, and neutron diffraction experiments do not show any trace of magnetic correlations. However, the presence of strong Sm$^{2+}$~-~Sm$^{2+}$ (ferromagnetic) exchange interactions already in the nearly divalent black phase of SmS has been demonstrated by ESR measurements \cite{BBR72} and an anomalous temperature dependence of the $^{11}$B relaxation rate in NMR experiments below 15~K points towards the presence of magnetic correlations in SmB$_{6}$ at ambient pressure too \cite{TYK81}. Our studies therefore seem to confirm the presence of slowly fluctuating moments in the semiconducting phase of SmB$_{6}$. Owing to the divalent-like character of the ground state wavefunction of SmB$_{6}$, the exchange interactions mentioned above can induce a strong mixing of the nonmagnetic ground multiplet $^{7}F_{0}$ of divalent Sm with its magnetic first excited multiplet $^{7}F_{1}$, which lies only $\sim$~36~meV ($\sim$~420~K) higher in energy. Moreover, the magnetic excitation spectrum of SmB$_{6}$, as measured by inelastic neutron scattering \cite{AMR95,MiA95}, shows the presence below $\sim$~50~-~100~K of a sharp transition at only 14~meV ($\sim$~160~K), which is interpreted as an excitation from the "true" intermediate valent ground state ($J$~=~0) into its first excited state ($J$~=~1). In this temperature range the $J$ mixing could therefore be even much stronger than expected for the free Sm$^{2+}$ ion, thus making the appearance of short-range magnetic correlations possible.

Recently a simple expression has been proposed \cite{Flo04a,Flo04b} for the extension to a lattice of the Kondo temperature ($T_{\mathrm{K}}$) formula for a single Ce impurity [$T_{\mathrm{K}}$~=~(1-$n_{f}$)$\Delta$, with $n_{f}$ and $\Delta$ the occupation number of the lower valent configuration (trivalent for Ce and divalent for Sm and Yb) and the width of the virtual bound state, respectively]. For a lattice, $\Delta$ can be replaced by the Fermi temperature [$T_{\mathrm{F}}$~=~$(1-n_{f})^{2/3}$$D_{\mathrm{0}}$]. In the case of Sm (where the equilibrium of the valence is given by the expression Sm$^{2+}$~$\leftrightarrow$~Sm$^{3+}$~+~5\textit{d} and the valence can be expressed as $v$~=~3~-~$n_{f}$), this gives $T_{\mathrm{KL}}$~=~$n_{f}$$(1-n_{f})^{2/3}$: this implies that $T_{\mathrm{KL}}$ will reach a maximum as $n_{f}$ goes from 0 to 1 (whereas it increases continuously in the case of Ce). When $k_{\mathrm{B}}$$T_{\mathrm{KL}}$ becomes lower than the crystal field splitting, slow relaxation processes are expected and thus magnetic ordering occurs.

\section{The effect of pressure on metallic Sm}

\begin{figure}[tbh]
\begin{center}
\includegraphics[scale=0.5,clip=]{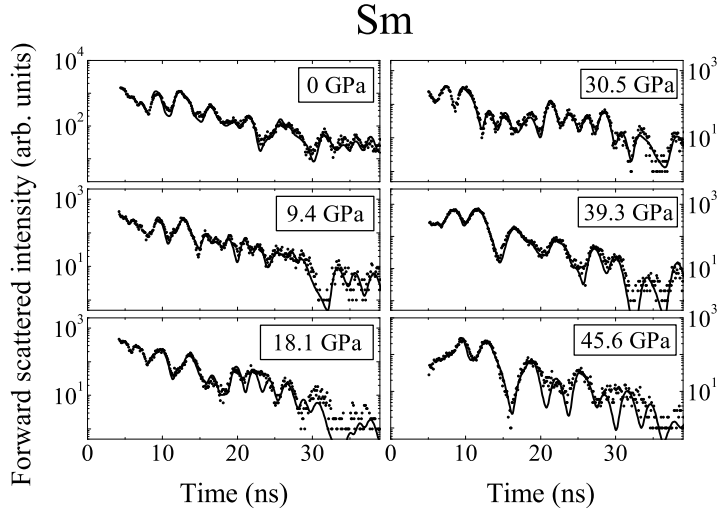}
\end{center}
\caption{\label{fig:six} $^{149}$Sm NFS spectra of metallic Sm at $T$~=~3~K for some selected pressures. The dots represent experimental data points, while the lines are fits.}
\end{figure}
\begin{figure}[tbh]
\begin{center}
\includegraphics[scale=0.5,clip=]{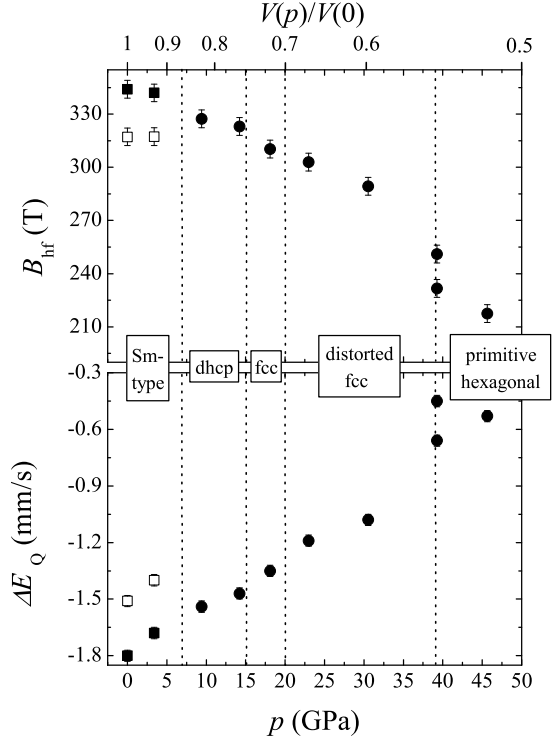}
\end{center}
\caption{\label{fig:seven} Dependence on pressure and reduced volume of the magnetic hyperfine field $B_{\mathrm{hf}}$ and of the electric quadrupole interaction $\Delta$$E_{\mathrm{Q}}$ of Sm metal at $T$~=~3~K. The symbols \fullsquare\ and \opensquare\ refer to the hexagonal and cubic sites of the Sm-type structure, respectively, while \fullcircle refer to the single site for the other structural phases. The reduced volume scale is approximately determined from reference~\cite{ZPH94}. Note the coexistence of the distorted fcc and primitive hexagonal phases at 39~GPa.}
\end{figure}

NFS measurements of metallic samarium have been performed at temperatures between 3 and 300~K in the pressure range 0~-~46~GPa, spanning over five different crystallographic structures. Typical measured spectra are shown in figure~\ref{fig:six}, while the pressure dependence of the saturation ($T$~=~3~K) hyperfine parameters is shown in figure~\ref{fig:seven}.

\subsection{Magnetic properties at ambient pressure} 

At ambient pressure, in the Sm-type structure, two inequivalent Sm sites with hexagonal or cubic coordination are present, in a ratio of 2:1, respectively. At high temperatures, above $T_{\mathrm{N}h}$~=~106~K, both sites are expected to show an electric quadrupole interaction due to the non cubic environment of all Sm atoms, owing to the fact that the ratio $c$/$a$ of the lattice parameters is smaller than its ideal value for the Sm-type structure. By comparison with the measured values of the electric field gradients in dhcp La metal \cite{PTW70}, one can roughly estimate the lattice contribution to the electric quadrupole interaction of both sites in Sm metal to be very small, of the order of only $\sim$~+~0.15~mm/s. Although the NFS spectra are not very well resolved at high temperatures, the best fit to the data gives $\Delta$$E_{\mathrm{Q}}^{h}$~=~0.0(1)~mm/s for the hexagonal sites and $\Delta$$E_{\mathrm{Q}}^{c}$~=~+0.2(1)~mm/s for the cubic sites at $T$~=~200~K. At low temperatures, below $T_{\mathrm{N}c}$~=~14~K, it is expected that both sites show a combination of magnetic hyperfine and electric quadrupole interactions. Here $|$$\Delta$$E_{\mathrm{Q}}$$|$ should be considerably enhanced with respect to its high temperature value, because a dominant 4\textit{f} term adds to the lattice contribution (which has ususally only a slight dependence on temperature). The analysis of the NFS spectra reveals that the magnetic hyperfine field for the hexagonal sites at $p$~=~0 and $T$~=~3~K, $B_{\mathrm{hf}}^{h}$~=~344(5)~T, is very close to the value expected for a free Sm$^{3+}$ ion (338(6)~T, reference~\cite{Ble72}), and it compares well with the average value found at 4.2~K by M\"ossbauer spectroscopy (345(18)~T) \cite{OfN67}. However, the electric quadrupole interaction $\Delta$$E_{\mathrm{Q}}^{h}$~=~-1.80(3)~mm/s is considerably reduced with respect to the free ion value of -2.1~mm/s \cite{Ble72} and the lattice contribution alone cannot account for the difference. Moon and Koehler (reference~\cite{MoK73}) have calculated approximate ground state wavefunctions for the cubic and hexagonal sites of Sm taking into account the crystal field and exchange interactions, in order to explain their measured neutron scattering amplitudes. For the hexagonal sites, in particular, their calculations reveal that the relatively strong exchange interactions, responsible for the high N\'eel temperature of 106~K, induce a mixing of the $J$~=~5/2 ground state with the $J$~=~7/2 excited state, with a reduction of the 4\textit{f} magnetic moment of these sites (0.56~$\mu_{\mathrm{B}}$) with respect to the Sm$^{3+}$ free ion value (0.71~$\mu_{\mathrm{B}}$). Using the same wavefunction, we have calculated the expected values of the magnetic hyperfine field and of the electric quadrupole interaction for this ground state, which amount to $\sim$~350~T and $\sim$~-~1.8~mm/s and are therefore in good agreement with our measured values. This is therefore one example of the complicated relationship between magnetic moment and hyperfine magnetic field which is quite typical for Sm compounds, where the mixing of multiplets with different $J$ can at the same time induce a reduction of the magnetic moment as compared to the free ion value and an increase of the magnetic hyperfine field.

At the cubic sites, the analysis of the NFS spectra indicates that the hyperfine parameters are considerably reduced with respect to those of the hexagonal sites: $B_{\mathrm{hf}}^{c}$~=~317(5)~T and $\Delta$$E_{\mathrm{Q}}^{c}$~=~-1.51(3)~mm/s, and even if the larger lattice contribution to the electric field gradient at the cubic sites (as determined at high temperature) is considered, the difference remains. A possible explanation for this can be given if one interprets the fact that the ordering temperature of the cubic sites ($T_{\mathrm{N}c}$~=~14~K) is far lower than that of the hexagonal sites ($T_{\mathrm{N}h}$~=~106~K) as a consequence of the reduced strength of the exchange interactions at the cubic sites, which can be reflected in a lower degree of mixing of the $J$~=~7/2 excited multiplet into the ground one. The calculation of Moon and Koehler (reference~\cite{MoK73}) shows indeed that the ground state wavefunction of the cubic sites does not contain any mixing with the $J$~=~7/2 multiplet at all and that a larger 4\textit{f} magnetic moment is expected for these sites. By using the wavefunction which best fits their neutron data, we obtain values of $\sim$~270~T for the magnetic hyperfine field and $\sim$~-~1.5~mm/s for the 4\textit{f} contribution to the electric quadrupole interaction. Although these values are lower than our measured ones, they qualitatively confirm that a reduction of the exchange interaction is in this case reflected in a decrease of the values of both magnetic hyperfine field and electric quadrupole interaction. We think that our set of data (including the complete temperature dependence of $B_{\mathrm{hf}}$ and $\Delta$$E_{\mathrm{Q}}$) in combination with the neutron data of Koehler and Moon (reference~\cite{KoM72}) might help evaluating better the effect of crystal field and exchange interactions on the ground state properties and especially on the value of the ordered magnetic moment of metallic Sm at ambient pressure, which is still a subject of controversy \cite{DDM99}.

\subsection{Magnetic moment decrease at high pressure}
As pressure increases, the Sm-type structure is expected to be stable up to $p$~$\sim$~7~GPa at 4.2~K. At $p$~=~3.4~GPa we observe indeed still the presence of hexagonal and cubic sites in the ratio 2:1, with different hyperfine interactions, but in both cases slightly reduced with respect to the ambient pressure values. Starting from $p$~=~9.4~GPa, in correspondence with the transition to the dhcp phase (La-type structure, stable up to $\sim$~15~GPa at 4.2~K), the NFS spectra can be properly analyzed with a single set of hyperfine parameters at $T$~=~3~K. Although in this structure two different crystallographic sites still exist for Sm, with hexagonal and cubic coordination, respectively, and in the ratio 1:1, they seem therefore to be characterized by the same values of their hyperfine parameters. This is in agreement with the resistivity studies of reference~\cite{DLD87}, where only one magnetic phase transition is observed for $p$~$>$~6~GPa. The dhcp structure of Sm metal can be stabilized at ambient pressure, if samples are produced in the form of thick films on appropriate substrates. Studies of these films with resonant magnetic x-ray scattering and neutron scattering \cite{DDS02} reveal that, despite the presence of two crystallographically inequivalent sites, Sm still orders antiferromagnetically, as in the Sm-type structure, but only one ordering temperature ($T_{\mathrm{N}}$~=~25~K) is observed for both hexagonal and cubic sites, owing probably to the enhanced coupling of layers with different coordinations and the consequent increase of the mutual exchange. Our results are in perfect agreement with these findings, and the hyperfine interactions we determine from our measurements at 9.4~GPa have an intermediate value between those of the two sites in the Sm-type phase (see figure~\ref{fig:seven}).

When pressure is increased further, the successive structural transitions from dhcp to fcc, from fcc to distorted fcc and from distorted fcc to primitive hexagonal are crossed at 15, 20 and 39~GPa, respectively \cite{ZPH94}. At 39~GPa the coexistence of the two phases is responsible for the presence of two sets of hyperfine parameters as shown in figure~\ref{fig:seven}, however, at 46~GPa the pure primitive hexagonal phase is observed. From figure~\ref{fig:seven} it is evident that at 3~K the strengths of both magnetic hyperfine field and electric quadrupole interaction decrease considerably in the pressure range between 0 and 46~GPa, reaching the values of $B_{\mathrm{hf}}$~=~218(5)~T and $\Delta$$E_{\mathrm{Q}}$~=~-0.53(3)~mm/s at the highest pressure, corresponding to relative reductions of $\sim$~40\% and $\sim$~70\% with respect to the ambient pressure values, respectively. In the same pressure interval, the volume of the unit cell shrinks by almost 50\% \cite{ZPH94}. The large reduction of the magnetic hyperfine field can be due to the variation with pressure of its different components. The largest contribution to the measured value is due to the 4\textit{f} electrons and is directly related to their magnetic moment, but other contributions arise from the polarization of core electrons and of conduction electrons. In Sm compounds the core polarization has the same sign as the 4\textit{f} contribution but a considerably smaller value (that can be estimated to amount to $\sim$~15~T), and it is generally considered as not dependent on pressure \cite{PKG93}. The conduction electron polarization can, on the other hand, give a large contribution to the hyperfine field, with possibly also a large pressure dependence. At the present stage it is therefore not possible to state clearly whether the strong reduction of $B_{\mathrm{hf}}$ is due to a corresponding reduction of the 4\textit{f} magnetic moment or to a simple change in the balance between the 4\textit{f} and the conduction electron contributions. However, the decrease of $B_{\mathrm{hf}}$ is associated to an even larger decrease of the electric quadrupole interaction $\Delta$$E_{\mathrm{Q}}$, which cannot be explained by the variation of its lattice component alone. We therefore believe that the large decrease of both $B_{\mathrm{hf}}$ and $\Delta$$E_{\mathrm{Q}}$ at the same time can be taken as an indication that the 4\textit{f} magnetic moment of Sm is reduced by pressure. Moreover, although the decrease of the hyperfine parameters is mostly continuous, the entrance into the primitive hexagonal high pressure phase at 39~GPa seems to be accompanied by a small discontinuity (downwards) of both $B_{\mathrm{hf}}$ and $\Delta$$E_{\mathrm{Q}}$, which can be interpreted as a further sign of the incipient delocalization of the 4\textit{f} electrons, which are proposed to be involved in the chemical bonding in this phase \cite{ZPH94}.

\section{Conclusions}

In conclusion we have studied the effect of external pressure on the magnetic and electronic properties of several Sm compounds, which are nearly divalent (SmS), intermediate valent (SmB$_{6}$) or trivalent (Sm metal) at ambient pressure. In both SmS and SmB$_{6}$, where the valence is known to increase smoothly with pressure, we observe the onset of long-range magnetic order at a critical pressure which corresponds approximately to the entrance of these systems into the metallic state. The ordered state is very stable towards pressure and has the properties expected for trivalent compounds, although both systems are still intermediate valent when the magnetic order appears. For pressures just below that for the onset of long-range order we observe the presence at low temperatures of short-range magnetic correlations of dynamical origin. These correlations are found to persist in SmB$_{6}$ even at ambient pressure. In the case of metallic Sm, pressure induces a strong decrease of the magnetic hyperfine field and the electric quadrupole interaction at the $^{149}$Sm nuclei. This can be related to a smooth decrease of the Sm 4\textit{f} moments when pressure increases. The small discontinuity in the pressure dependence of the hyperfine field when Sm undergoes the transition into the primitive hexagonal phase at 39~GPa might indicate the onset of the delocalization of the 4\textit{f} electrons. However, further studies are necessary at higher pressures in order to obtain the full delocalization.

\end{document}